# Investigating Wake Structures in Flow Past Oscillating Cylinder Using Proper Orthogonal Decomposition

Anubhav Sinha[1*], and Parasuram IVLN[2]

[1]Department of Mechanical Engineering, Indian Institute of Technology (BHU), Varanasi-221005, India
[2]Department of Mechanics, KTH Royal Institute of Technology, Stockholm, Sweden

## Abstract

*The present study investigates unsteady wake dynamics in flow past oscillating cylinder. Two-dimensional computational study is carried out for flow past sinusoidally oscillating cylinder. Operating parameters like Reynolds number, forcing frequency, amplitude and wavelength are systematically varied and their effect on wake dynamics is investigated. Lift and drag coefficients are plotted with cylinder position to characterize the flow. Further, flow field is investigated using Proper Orthogonal Decomposition (POD). Time-resolved vorticity contours are obtained. These results are then processed using POD algorithm. POD mode shapes and Power Spectral Density (PSD) plots are obtained for each case. Wake dynamics is analyzed using POD results. Flow response to cylinder oscillations is characterized by the POD mode shape and dominant frequency peaks in PSD plots. It is observed that flow is too dissipative at lower values of Reynolds number. Further, effect of frequency ratio is observed in achieving synchronized vortex shedding. High amplitude and large wavelength regime, which remained unexplored in previous research is also investigated.*

\* Corresponding author email – er.anubhav@gmail.com

# Introduction

Vortex induced vibrations (VIV) is a topic of much interest owing to its practical applications in structures like chimney stack, bridge, heat exchanger tubes, etc. The rich physics behind unsteady vortex shedding have attracted fundamental research investigations since decades (Bearman, 1984; Williamson and Govardhan, 2004, 2008). Usually, an elastic cylinder, or elastically tethered cylinder in a uniform stream is studied. Force measurements and flow visualizations are carried out to gain understanding of this configuration. However, a more controlled experiment can be designed where the cylinder follows a sinusoidal forcing function transverse to a uniform flow. This configuration is explored in the present study. The emphasis is on identifying important flow features, dominant frequencies, and assessing the effectiveness of Proper Orthogonal Decomposition (POD) to characterize such flow fields.

Previous research on this subject have been summarized in various reviews by Sarpkaya (1979), Bearman (1984), Griffin & Ramberg (1982), Williamson and Govardhan (2004, 2008). Williamson and Roshko (1988) experimentally investigated wake structures from flow past oscillating cylinder in a water tank. They have visualized vortex formation and have demonstrated that the oscillating cylinder produces four regions of vorticity each cycle. For smaller wavelengths, these regions coalesce and form a Karman-type wake. For larger wavelengths, these vortices convect away from each other. They carried out experiments on a wide range of conditions, varying amplitude ($0.2 < A/D < 1.8$), and wavelength ($1 < \lambda/D < 10$). Amplitudes and wavelengths are generally non-dimensionalized by the cylinder diameter ($D$) in literature. They limited their experiments to a Reynolds number ($Re$) of 392. Based on their visualization and interpretation of vortical structures, they proposed a regime map in the amplitude-wavelength plane. The regime map postulated that new patterns may emerge at higher wavelength and amplitude regions. This aspect will be investigated in the present study. Blackburn and Henderson (1999) have carried out numerical simulations to study two-dimensional flow past an oscillating cylinder. They maintaied the Reynolds number and oscillation amplitude constant in their study (Re=500, and A/D=0.25). They focus on the oscillation frequency of the cylinder and investigate the variation of λ/D near the natural shedding frequency (or Karman frequency) of stationary cylinder. A change in sign of mechanical energy transfer between fluid and cylinder is observed with variation in forcing frequencies. The forcing or excitation frequency ($F_e$) is non-dimensionalized using the natural shedding frequency of stationary cylinder ($F_n$) to obtain frequency ratio ($F$). Energy transfer from fluid to cylinder is observed for F=0.875 and from cylinder to the fluid for F=0.975. This aspect is explored in detail the present study.

Carberry and Sheridan (2001, 2003) experimentally investigated flow past an oscillating cylinder in a water tank. They have measured forces acting on cylinder and also carried out flow visualizations to understand wake formation. They have varied their excitation frequency from 05 to 1.4 times the Karman frequency. Reynolds number is held constant to 2300, and *A/D* has values 0.5 and 0.6. They report transition in wake structure while increasing the forcing frequency which strongly impacts lift force. Guilmineau, and Queutey (2002) carried out numerical study of an oscillating cylinder in a uniform flow. They have analyzed different vortex shedding patterns for various forcing frequencies at a Reynolds number of 185. Kim et al. (2014) studied turbulent flow past an oscillating cylinder for Reynolds number of 5500 using Large Eddy Simulations. For their study, A/D=0.5 and 0.75<F<0.95. They investigated vortex shedding using vorticity contours and report a transition between different modes between 2P and 2S modes. Ongoren and Rockwell (1988 a, b) have experimentally investigated flow past an oscillating cylinder. Reynolds number is varied between 584 to 1300. They have classified the observed vortex modes into symmetric and anti-symmetric. Lu and Dalton (1996) numerically examined flow past an oscillating cylinder. They observe shifting of high vorticity regions towards the cylinder with increase in forcing frequencies. However, after a critical frequency the trend reverses. This process is investigated in detail analyzing the vortex structure contours. Their simulations cover Reynolds number of 185, 500, and 1000. F varies from 0,8 to 1.2 and (A/D) has values of 0.4 and 1.

Previous research on flow past oscillating cylinder relied on flow visualization and force measurement to investigate this configuration. However, the unsteady dynamics of vortical structures, their spatial and temporal characterization are far too complex be investigated by instantaneous flow visualization only. Specialized tools are needed which not only provide information about the coherent structures but characterize the temporal response of wake structures. Proper Orthogonal Decomposition (POD) is a suitable tool for such requirements.

Riches et al. (2018) have undertaken experimental study of vortex induced vibrations of an elastic cylinder. They capture wake structures using Particle Image Velocimetry (PIV). The PIV images are processed using POD analysis to characterize the flow behavior. Konstantinidis et al. (2007) have presented experimental results of flow past oscillating cylinder for high frequency ratios (1.74< F <1.87) at Reynolds number of 2100-2200. They probe the bimodal switching of vortex between 2S and 2P modes for turbulent conditions. Zhang and Zheng (2018) have established the applicability of POD for flow past two tandem cylinders. The downwind cylinder is also oscillating transverse to the flow. They have investigated the temporal evolution of spatial modes using Lyapunov exponents of the temporal coefficients of the POD modes. Yao and Jaiman (2017) study vortex induced vibrations of a tethered cylinder for low Reynolds number. POD is used to analyze the simulations results produced in this study. Nabatian and Mureithi (2015) undertook numerical investigation of flow past oscillating cylinder, at Re=200, A/D=0.05-0.5, and frequency ratios 1, 1.5 and 2. They probe the symmetric and non-symmetric mode by carrying out POD analysis of the velocity field.

To summarize, wake dynamics of flow past oscillating cylinder has been the subject to several experimental and numerical investigations. However, most studies are typically carried out in a water tank with flow visualization or PIV. Forces on cylinder are measured to provide insight on fluid-structure interaction. Moreover, advanced tools like POD have rarely been used for flow past oscillating cylinders. The work of Nabatian and Mureithi (2015) appears to be the only work on application of POD to oscillating cylinder wake dynamics. However, that work is focused on much higher frequency ratios. Moreover, the regime map proposed by Williamson and Roshko (1988) is widely accepted and many of their experiments are reproduced by some researchers. However, very little information is available on the temporal characteristics of wake dynamics. Only instantaneous images are captured by previous researchers. There is still gap in areas of frequency response of vortical structures, especially near the critical frequency ratios. This paper attempts to address this issue. In the next section, computational modelling approach is explained, with introductory background on POD algorithm. List of cases and various conditions covered are listed in Table 2. Further, computational results and POD analysis is presented and discussed. The paper concludes with the summary on the results and observations.

# Computational Modelling

The study utilizes Ansys Fluent (Ansys, 2015) platform (version 2021 R2) to carry out two-dimensional simulations of flow past oscillating cylinder. Reynolds number is defined based on the cylinder diameter and flow velocity. As the Reynolds number is limited to 500 and lower values, it is reasonable to assume a laminar flow. Unsteady, pressure-based solver is employed for the simulations. SIMPLE algorithm (Patankar, 1980) is used. The computational domain, along with boundary conditions is shown in Fig. 1(a). The domain extends to 60 D along the flow direction and 30 D in the transverse direction, where D is the cylinder diameter. To establish grid-independence, three meshes are used. Number of cells in each mesh is given in Table 1. Typical results for lift coefficient from these meshes are shown in Fig. 1(b). As evident, results for all the three cases almost coincide. However, mesh M2 is used for further study, and is shown in Fig. 1(c).

| **Mesh** | M1 | M2 | M3 |
|---|---|---|---|
| **No of cells** | 27614 | 38827 | 59763 |

**Table 1. Mesh used for assessing grid-independence**

The cylinder is given sinusoidal forcing using function:

$$y(t) = A \sin \omega t \qquad (1)$$

Where y(t) is the cylinder displacement at time t, A is the amplitude of oscillation, $\omega$ is the angular frequency. $A$ and $\omega$ are varied to obtain various flow conditions. Fluid velocity is also varied to observe effect of change in Reynolds number. Case details are discussed in the next section and all parameters are listed in Table 2. Karman frequency or natural frequency is obtained by simulation of flow past stationary cylinders for various Reynolds number. Then simulations with oscillating cylinders are performed. Lift and Drag values are obtained. Velocity field is processed to obtain vorticity plots. These plots are used in POD algorithm to investigate wake dynamics. Mathematical background, applications and uses of POD can be found in a seminal paper by Berkooz et al. (1993). Recent reviews by Taira et al. (2017, 2020) provide useful overview of various decomposition techniques available, basics of POD and its applications. The POD algorithm used in this study is validated in a previous experimental study (Sinha and Ravikrishna, 2019). POD is a data reduction technique which provides insight on temporal and spatial modes of a flow field. Imagine the parameter under investigation to be a function of space and time, which can be represented as:

$$Z(x,t) = \sum \alpha_k(x)\beta_k(t) \qquad (2)$$

For example, in this study, values of vorticity is captured in a 2D plane which give the spatial information on vorticity. Data for different time steps are saved, which give temporal information on vorticity. The total data set can be assumed to represent Z. Now the challenge is that functions $\alpha$ and $\beta$ are not unique and will depend on selection of basis functions. POD can be understood as the method to obtain optimal functions $\alpha$ and $\beta$. POD aims to find orthonormal basic functions such that the minimum number of modes are required to represent function Z. To achieve this objective, modes are arranged in order of decreasing modal energy. Hence, the first *n* modes gives the best possible *n*-term representation of the function Z. $\alpha$ gives the spatial information and will be represented as POD mode shapes. $\beta$ retains the temporal information. Fast Fourier Transform (FFT) of $\beta$ will provide Power Spectral Density (PSD) plots, which show dominant frequencies associated with that mode.

# Results and Discussion

Table 2 lists the various cases investigated in this study and relevant parameters. The cases can be divided into three sets. First set comprises of cases C1 to C3. In this set, for a fixed ($\lambda/D$), effect of $Re$ and ($A/D$) are analyzed. The second set comprise of cases C4 to C7. For this set, the excitation frequency is near the natural frequency of a stationary cylinder. For these cases, the effect of frequency ratio ($F$) and Reynolds number is under investigation. The last set consists of cases C8 to C11. These cases explore the large ($A/D$) and ($\lambda/D$) regions which might exhibit new patterns as pointed out by Williamson and Roshko (1988). Effect of Reynolds number is also examined for this set. In the following paragraphs, these sets of simulation cases are examined using $C_L$ and $C_D$ plots, POD mode shapes and PSD plots.

| Case | Amplitude (A/D) | Wavelength ($\lambda$/D) | Reynolds Number ($Re$) | Excitation/ forcing frequency ($F_e$) | Natural frequency ($F_n$) | Frequency ratio $\left(F = \dfrac{F_e}{F_n}\right)$ |
|---|---|---|---|---|---|---|
| C1  | 0.40 | 5.00  | 100 | 0.1205 | 0.100 | 1.2046 |
| C2  | 0.40 | 5.00  | 200 | 0.2409 | 0.220 | 1.0951 |
| C3  | 1.20 | 5.00  | 200 | 0.2500 | 0.220 | 1.1364 |
| C4  | 0.25 | 6.26  | 200 | 0.1925 | 0.220 | 0.8750 |
| C5  | 0.25 | 5.62  | 200 | 0.2145 | 0.220 | 0.9750 |
| C6  | 0.25 | 6.01  | 500 | 0.5014 | 0.573 | 0.8750 |
| C7  | 0.25 | 5.39  | 500 | 0.5587 | 0.573 | 0.9750 |
| C8  | 1.50 | 30.0  | 200 | 0.0402 | 0.220 | 0.1825 |
| C9  | 1.50 | 30.0  | 500 | 0.1004 | 0.573 | 0.1752 |
| C10 | 1.20 | 17.0  | 200 | 0.0709 | 0.220 | 0.3221 |
| C11 | 1.20 | 17.0  | 500 | 0.1772 | 0.573 | 0.3092 |

**Table 2. Details of cases investigated in this study**

POD mode shapes capture the dominant coherent structures in the flow field. The first POD mode denotes the mean flow field. PSD spectrum corresponding to first mode is not shown since it only exhibits zero frequency (as expected for a mean mode). Second and third POD modes are almost identical. They denote wake structures moving past the cylinder. The behavior is similar to what was observed for the second and third POD modes for a stationary cylinder by Taira et al. (2020). Further modes exhibit small-scale structures with lower energy content. PSD plots are usually shown based on frequency. However, for this study, normalized frequency is used where the observed frequency is normalized by the excitation frequency. For plotting $C_L$ and $C_D$ variations, at least 20 cycles of data is used for each case.

Figure 2 shows the variation of $C_L$ and $C_D$ with $y/D$ for cases C1-C3. These cases have the same wavelength ($\lambda/D = 5$). C1 and C2 are at different Reynolds number while C2 and C3 have different amplitude ratios. For C1 and C2, $C_L$ remains approximately a linear function of $y/D$. Whereas $C_L$ becomes a more complex function in C3, which has higher amplitude. Coefficient of drag ($C_D$) is a function of ($y/D$) location and phase of the cylinder. For C1 and C3, it is close to phase locked condition, while for C2, it exhibits more chaotic behavior, with significant cycle-to-cycle variations. The cycle-to-cycle variations are maximum in C2 and minimum in for the high amplitude case C3.

POD mode shapes are depicted in Fig. 3, and corresponding PSD spectra are shown in Fig. 4. Mode 1 of C1 and C2 show the typical behavior of a low amplitude, low $Re$ case. Two streams of vorticity are getting formed at the upper-most and lower-most position of the cylinder. C1 and C2 cases correspond to the 2S pattern as observed by Williamson and Roshko (1988). Mode 2 and 3 show the structures formed by the cylinder wake. C2 matches the pattern of the 2S configuration closely. However, wake in C1 case appear different. A close examination reveals that the vortex formed from cylinder oscillations in C1 gets dissipated rather quickly; and the characteristics of flow past stationary cylinder (Kaman vortex) dominates. This behavior gets manifested as the elongated structure observed in mode 2 and 3. It is relevant to note that the PSD spectra is also bimodal where $F_e$ is not the dominant peak. This could be attributed to the highly viscous nature of flow at $Re=100$. For C1, mode 4 and higher modes show small-scale structures which do not contain significant energy. PSD spectrum of 2nd and 3rd modes of C1 show a dominant peak of around 0.6 times its excitation frequency ($F_e$). There is also a minor peak at $F_e$. For higher modes, the dominant peaks are at the $F_e$, while there is a minor peak at around 0.6 $F_e$. Mode 6 appears to show large scale oscillations, while its dominant peak is much smaller than $F_e$.

C2 results are very different from that of C1. For the 2nd and 3rd modes, the dominant peak is at $F_e$ and minor peaks for other frequencies. This can be attributed to the combined effect of cylinder oscillations and uniform flow past the cylinder. The PSD spectrum is also broader suggesting the presence of different frequencies, and the flow not being dissipative like C1. This also explains the large cycle-to-cycle variation as observed in $C_D$ plot in Fig. 2. POD mode shapes for 2nd and 3rd mode corresponds to the 2S classification of Williamson and Roshko (1988). Higher modes (4 and 5) show small-scale structures with lower energy content. Case C3 has the same Re as C2 (Re=200), and same $\lambda/D$, however, the amplitude is much higher than C1 or C2 (A/D=1.2). POD mode 1 shows the mean flow field. Mode 2 and 3 show the vortex pattern in the wake region. PSD plots for mode 2 and 3 show that they correspond to $F_e$. PSD plots reveal single dominant peak at $F_e$, which results in synchronized wake formation, as evident in $C_L$ and $C_D$ plots shown in Fig. 2. Higher modes show small-scale structures which have dominant frequency at $2F_e$ and $3F_e$. It is to be noted that case C3 corresponds to the (P+S) region in the regime map proposed by Williamson and Roshko (1988). It is encouraging to note that these two patterns are captured clearly using POD and PSD plots. With the above discussion, it can be concluded that $Re=100$ is too dissipative to observe the effect of cylinder oscillations in wake patterns. It also explains why previous research is focused on higher Reynolds number regimes. Re=200 seem acceptable for now. But it will also be examined in the next set of cases. It is further observed that higher amplitudes (A/D=1.2) results in lower cycle-to-cycle variations. However, as will be shown in the next few paragraphs, even for lower amplitudes (A/D=0.25), phase locking can be observed, provided the frequency ratio is in the suitable range.

The next set of cases explore the role of frequency ratio (*F*) and Reynolds number (*Re*). Case parameters are selected based on the study of Blackburn and Henderson (1999). Their simulations are carried out at low amplitudes (A/D=0.25) and are limited to Re=500. Blackburn and Henderson (1988) reported change in sign of mechanical energy transfer as *F* is changed from 0.875 to 0.975. This behavior needs to be examined in detail owing to its importance in practical applications. This will be investigated with the aid of POD analysis. First, consider cases C4 and C5. They both have Re=200. F=0.875 for C4 and F=0.975 for C5. $C_L$ and $C_D$ plots for cases C4-C7 are shown in Fig. 5. $C_L$ and $C_D$ plots show perfect phase locking for this set of cases. It is to be noted that the data used to make plots in Fig. 5 consists of at least 20 cycles, and still overlap on a thin curve demonstrating almost zero cycle-to-cycle variations. Compare this to the scatter observed in Fig. 2 for case C2. It is interesting to note that C2 has same Re as and very similar ($\lambda$/D) to case C5 (cf. Table 2). Only difference lies with the *F* values. *F* is higher than 1 for C2 and lower than 1 for C5. This observation demonstrates the significance of *F* and its impact on cylinder wake dynamics.

POD mode shapes for C4 and C5 are shown in Fig. 6. PSD plots for these cases are shown in Fig. 7. The PSD plots reveal the reason for synchronization observed in Fig. 5. The dominant peaks are all at the excitation frequency ($F_e$) for 2nd and 3rd modes and $2F_e$ for 4th and 5th mode. Sixth mode however shows dominant peak at $3F_e$. Similar behavior is observed for both C4 and C5, except for the fact that C5 has broader peaks. POD mode 1 shows the mean flow field. Modes 2 and 3 show the vortical structures in the wake of the oscillating cylinder. For the given parameters, cases C4-C7 will fall in the 2S region of regime map of Williamson and Roshko (1988). The wake structures observed in mode 2 and 3 of C4 and C5 match well with the 2S pattern. It is important to point that POD modes for C4 and C5 cases look almost identical. Hence, the switch in behavior as reported by Blackburn and Henderson (1988) is not evident for these cases (from the POD modes). This again can be attributed to the dissipative nature of low *Re* flow (even for *Re*=200).

For cases C6 and C7, POD mode shapes are shown in Fig. 8 and PSD plots in Fig. 9. Mode 1 shows the average flow field. It is similar to the mean flow field in the previous pair C4 and C5. However, features are more prominent for C6 and C7 as compared to C4 and C5. Mode 2 and 3 characterizing the wake structures highlight a major difference between C6 and C7. As evident, wake structures are markedly larger for C6 as compares to C7. Why this difference should be there even though all parameters are similar, and F is also close (being 0.875 for C6 and 0.975 for C7)? This can be explained on the basic of observations reported by Blackburn and Henderson (1999). They investigated cases C6 and C7 and have reported phase switch in transfer of mechanical energy. They observed that there is a net transfer of energy from the fluid to the cylinder for C6 conditions; and from cylinder to the fluid for C7 conditions. Wake structure observed in mode 2 and 3 characterize the deficiency in fluid momentum. Hence, case where energy is extracted from the fluid (case C6) wake is expected to be larger. Case where energy is transferred to the fluid (case C7), wake will be smaller. This is indeed the case with wake structures observed in Fig. 8. Further, PSD plots show a sharp peak at the forcing frequency ($F_e$) for modes 2 and 3; at $2F_e$ for mods 4 and 5; and at $3F_e$ for the 6th mode. The peaks are also narrow highlighting the synchronized behavior of these wakes. The above discussion also confirms the utility of POD as a tool to identify phase switch in energy transfer. Results from this set of cases also confirm that a minimum value of Reynolds number is required to overcome dissipation and for these processes to be observed.

The last set of cases (C8-C11) report unexplored regimes in previous research. Williamson and Roshko (1988) have expressed the possibility of new pattens for high amplitudes and large wavelengths values (cf. Fig. 3 of Williamson and Roshko, 1988). Based on these guidelines, parameters for cases C9-C11 are selected. Fig. 10 presents the $C_L$ and $C_D$ plots for cases C8-C11. It is observed that lift and drag in these cases are very different from previous cases. They are not phase-locked like the previous set (C4-C7), and large cycle-to-cycle variation is evident. However, this variation appears to have some order and is not random or chaotic. More insight is gained by POD modes and PSD spectra. POD mode shapes and PSD plots for cases C8 and C9 are shown in Fig. 11 and 12. POD mode 1 represents the average flow field behavior. Mode 1 features of C9 are more prominent as compared to C8. Further, mode 2 and 3 of C9 appear to capture the vortex shedding accurately. These modes also show a dominant peak at the forcing frequency. Mode 4 shows the higher harmonic at $2\ F_e$, while mode 4 and 5 show small-scale structures at much higher frequencies. For case C8, only mode 2 is at the forcing frequency. For case C8, mode 2 shows similar features like mode 2 of C9, and also PSD peak at $F_e$. But the features appear to be faint in this mode. A more accurate representation of cylinder wake is shown in mode 5, which closely represents mode 2 of C9. This is also confirmed by PSD plots which show a peak at $F_e$ for this mode. Other modes for this case show smaller structures and frequencies different than $F_e$. The fact that the mode which captures most important feature is 5th mode, and modes with small-scale features take precedence highlights the dissipative nature of flow at Re=200.

POD mode shapes for C10 and C11 are shown in Fig. 13 and corresponding PSD plots are shown in Fig. 14. Case C11 demonstrate the characteristics of a phase-locked situation. The first mode prominently shows the average flow domain. Second and third modes depict the cylinder vortex patterns which also show dominant peak at the excitation frequency. These modes are also comparable with the corresponding modes of C9. Higher modes of C11 show smaller structures with higher frequency. It is interesting to note that the second mode of C11 has a small peak around 1.4 $F_e$, which is a minor peak in mode 2, but keep getting stronger in subsequent mode, and is the dominant peak at modes 4 and 5. For case C10, its POD modes are similar and less prominent than C11. Second and third mode also correspond to forcing frequency. However, they also have other frequency contributions. Higher modes tend towards higher frequency and small-scale structures.

It is noteworthy that the cases with higher amplitudes and ($\lambda/D$) have prominent flow features even for higher modes. While other cases had prominent features only up to 3-4 modes and modes 5 and 6 were mostly faint disturbances near the oscillating cylinder. This is also true for the high amplitude case C3. Cases having prominent features in higher modes can be understood to have higher energy in their unsteady behavior. And it appears that both high amplitude and high ($\lambda/D$) are required for it.

## Conclusions

The present work explores flow past oscillating cylinder by using two-dimensional computational modelling. Lift and Drag coefficients are obtained and compared. POD analysis based on vorticity contours is used to obtain mode shapes and dominant frequencies. First set of case demonstrates the significance of Reynolds number and amplitude of cylinder oscillation. It is observed that the POD modes are more prominent for Re of 200 and 500, and the vortex street becomes more prominent with increase in amplitude of oscillations. Second set of cases assess the effect of Reynolds number and frequency ratio. It is evident that the best response is obtained from cases with Re=500. This observation partly explains why relatively higher Re is used in the study of Blackburn and Henderson, R. D. (1999). Switch in direction of energy transfer is observed when F is changed from 0.875 to 0.975. POD mode shapes are able to distinguish between these two cases, even though the difference in frequency is not so large. This demonstrates the applicability of POD for this configuration. Further PSD plots are also used to characterize the temporal response of the wake. The last set investigates the unexplored conditions highlighted by Williamson and Roshko (1988). High amplitude and large wavelength regions of this map are explored. Coherent structures are formed and captured by POD modes.

## Acknowledgements

The authors gratefully acknowledge the support provided by Science and Engineering Research Board (SERB), Government of India through their Start-up Research Grant scheme.

# References

ANSYS Inc, ANSYS Fluent Theory Guide 12.0, 2015

Blackburn, H. M., & Henderson, R. D. (1999). A study of two-dimensional flow past an oscillating cylinder. *Journal of Fluid Mechanics*, *385*, 255-286.

Bearman, P. W. (1984). Vortex shedding from oscillating bluff bodies. Annual review of fluid mechanics, 16(1), 195-222.

Berkooz, G., Holmes, P., & Lumley, J. L. (1993). The proper orthogonal decomposition in the analysis of turbulent flows. Annual review of fluid mechanics, 25(1), 539-575.

Carberry, J., Sheridan, J., & Rockwell, D. (2001). Forces and wake modes of an oscillating cylinder. Journal of Fluids and Structures, 15(3-4), 523-532.

Carberry, J., Sheridan, J., & Rockwell, D. (2003). Controlled oscillations of a cylinder: a new wake state. Journal of Fluids and Structures, 17(2), 337-343.

Griffin, O. M., & Ramberg, S. E. (1974). The vortex-street wakes of vibrating cylinders. Journal of Fluid Mechanics, 66(3), 553-576.

Guilmineau, E., & Queutey, P. (2002). A numerical simulation of vortex shedding from an oscillating circular cylinder. Journal of Fluids and Structures, 16(6), 773-794.

Kim, S., Wilson, P. A., & Chen, Z. M. (2014). Numerical simulation of force and wake mode of an oscillating cylinder. Journal of Fluids and Structures, 44, 216-225.

Konstantinidis, E., Balabani, S., & Yianneskis, M. (2007). Bimodal vortex shedding in a perturbed cylinder wake. Physics of Fluids, 19(1), 011701.

Lu, X. Y., & Dalton, C. (1996). Calculation of the timing of vortex formation from an oscillating cylinder. Journal of Fluids and Structures, 10(5), 527-541.

Nabatian, N., & Mureithi, N. W. (2015). Lock-On Vortex Shedding Patterns and Bifurcation Analysis of the Forced Streamwise Oscillation of the Cylinder Wake. International Journal of Bifurcation and Chaos, 25(09), 1530022.

Ongoren, A., & Rockwell, D. (1988)a. Flow structure from an oscillating cylinder Part 2. Mode competition in the near wake. Journal of Fluid Mechanics, 191, 225-245.

Ongoren, A., & Rockwell, D. (1988)b. Flow structure from an oscillating cylinder Part 1. Mechanisms of phase shift and recovery in the near wake. Journal of fluid Mechanics, 191, 197-223.

Patankar, S. V. (1980). Numerical heat transfer and fluid flow, Hemisphere Publ. Corp., New York, 58.

Riches, G., Martinuzzi, R., & Morton, C. (2018). Proper orthogonal decomposition analysis of a circular cylinder undergoing vortex-induced vibrations. Physics of Fluids, 30(10), 105103.

Sarpkaya, T. (1979). Vortex-induced oscillations: a selective review.

Sinha, A., & Ravikrishna, R. V. (2019). Experimental studies on structure of airblast spray in crossflow. Sādhanā, 44(5), 1-13.

Taira, K., Brunton, S. L., Dawson, S. T., Rowley, C. W., Colonius, T., McKeon, B. J., ... & Ukeiley, L. S. (2017). Modal analysis of fluid flows: An overview. AIAA Journal, 55(12), 4013-4041.

Taira, K., Hemati, M. S., Brunton, S. L., Sun, Y., Duraisamy, K., Bagheri, S., ... & Yeh, C. A. (2020). Modal analysis of fluid flows: Applications and outlook. AIAA journal, 58(3), 998-1022.

Williamson, C. H., & Roshko, A. (1988). Vortex formation in the wake of an oscillating cylinder. *Journal of fluids and structures*, *2*(4), 355-381.

Williamson, C. H., & Govardhan, R. (2004). Vortex-induced vibrations. Annu. Rev. Fluid Mech., 36, 413-455.

Williamson, C. H. K., & Govardhan, R. (2008). A brief review of recent results in vortex-induced vibrations. Journal of Wind engineering and industrial Aerodynamics, 96(6-7), 713-735.

Yao, W., & Jaiman, R. K. (2017). Model reduction and mechanism for the vortex-induced vibrations of bluff bodies. Journal of Fluid Mechanics, 827, 357-393.

Zhang, M., & Zheng, Z. C. (2018). Relations of POD modes and Lyapunov exponents to the nonlinear dynamic states in flow over oscillating tandem cylinders. Physics of fluids, 30(12), 123602.

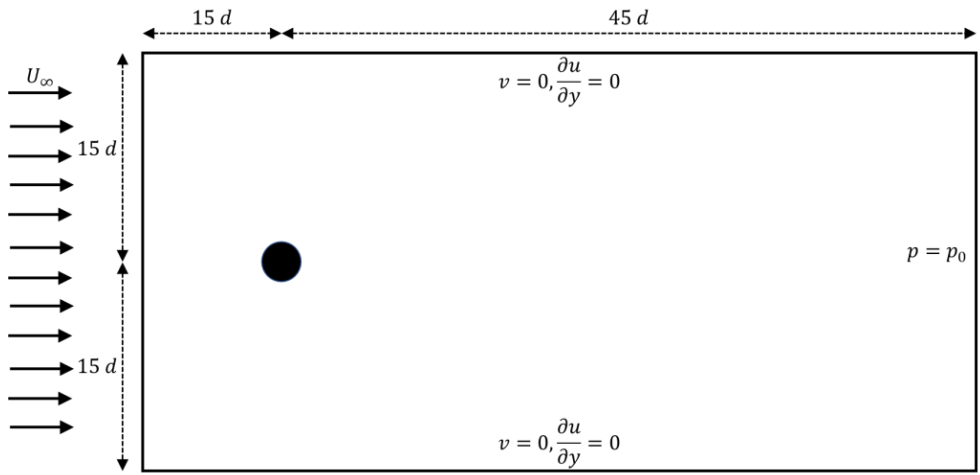

(a) Schematic of Computational Domain

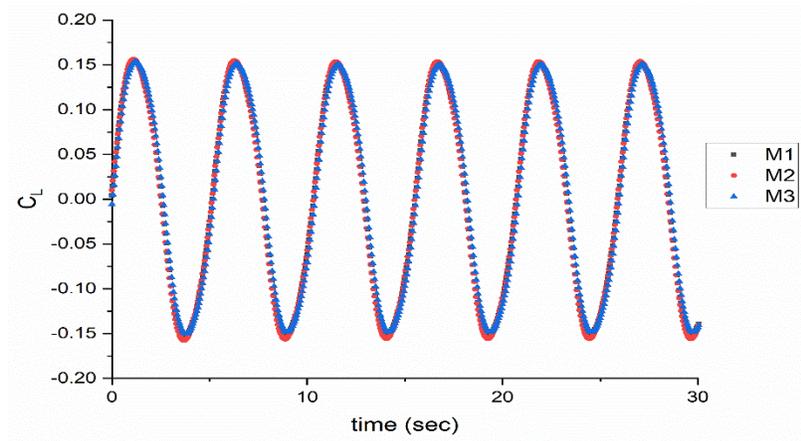

(b) Grid independence-coefficient of lift from simulations from M1, M2 and M3.

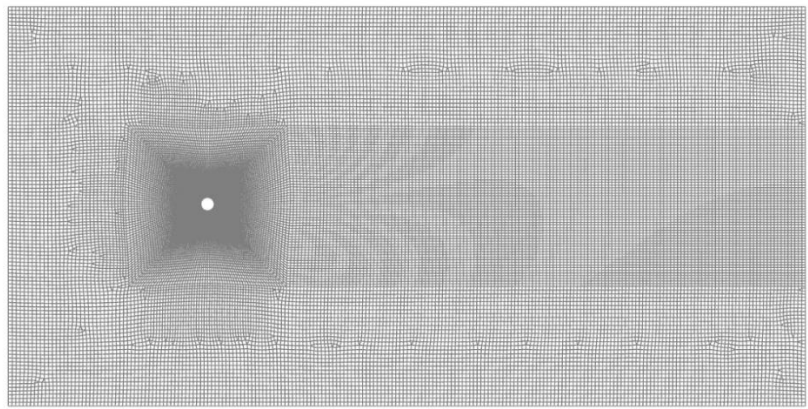

(b) Mesh used in the present study (M2)

**Figure 1. Details of the computational domain (a) Schematic and (b) Grid independence, (c) Mesh used for simulations (M2)**

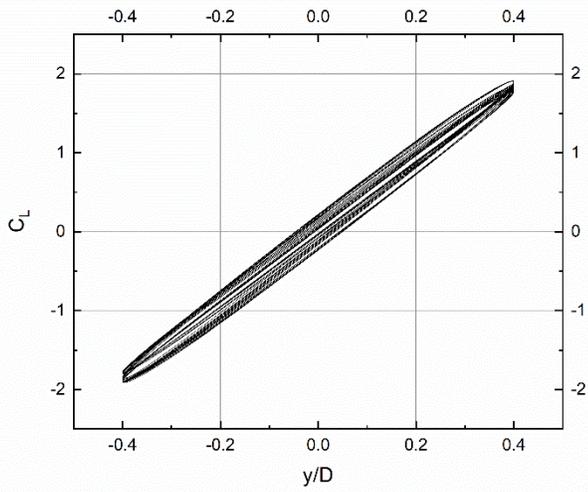
C1, lift coefficient

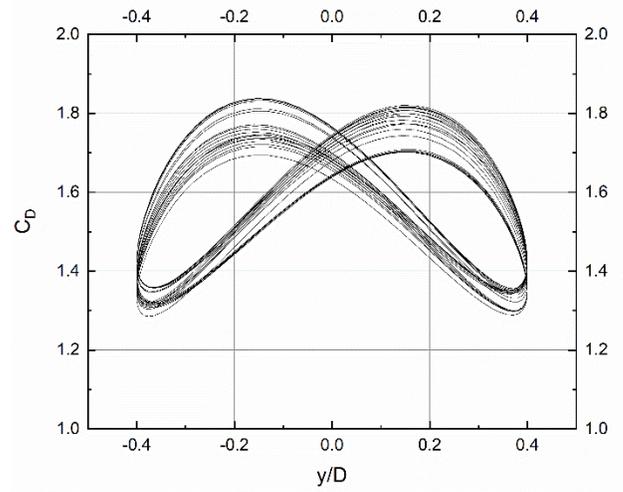
C1, drag coefficient

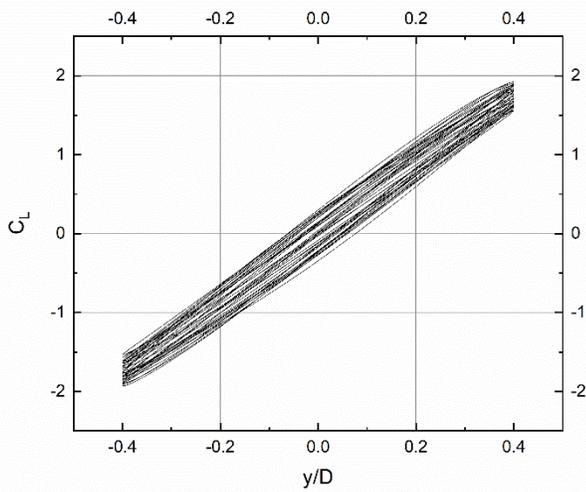
C2, lift coefficient

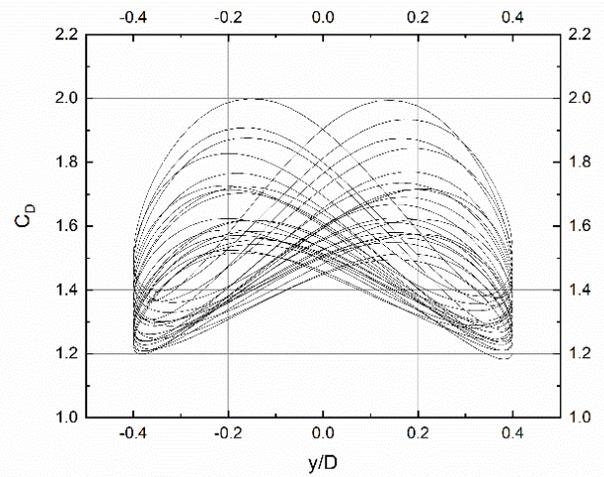
C2, drag coefficient

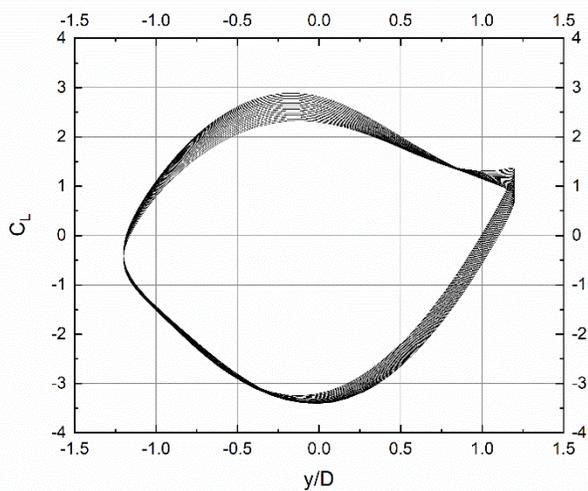
C3, lift coefficient

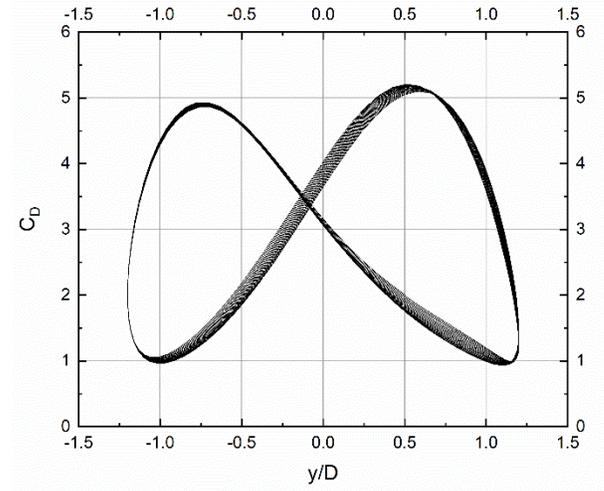
C3, drag coefficient

**Figure 2. Lift and Drag coefficients for cases C1, C2 and C3**

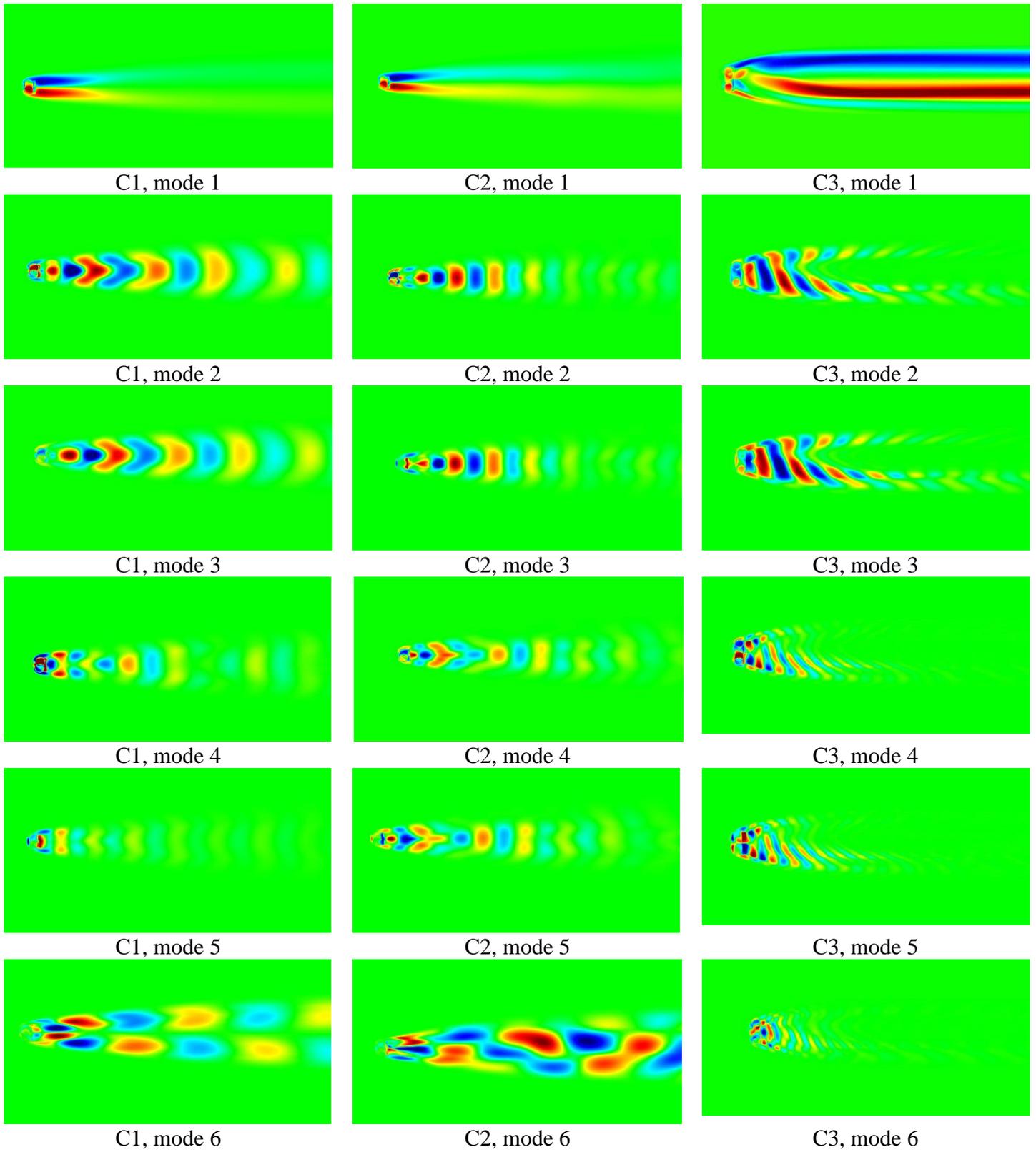

**Figure 3. POD modes for Cases C1, C2 and C3**

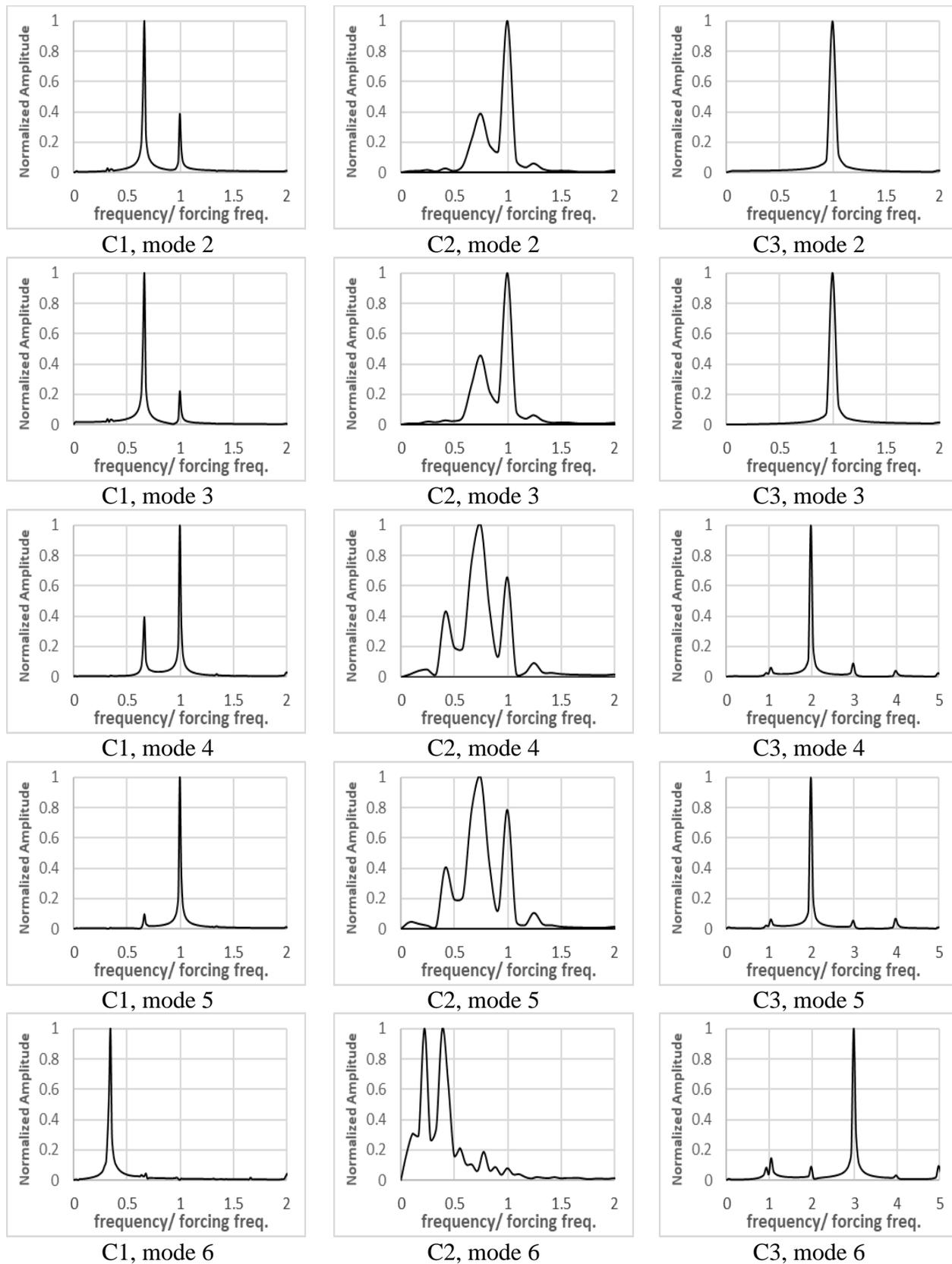

**Figure. 4. PSD plots for Cases C1, C2 and C3**

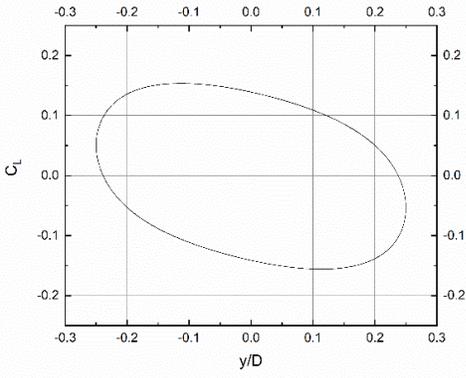
C4, lift coefficient

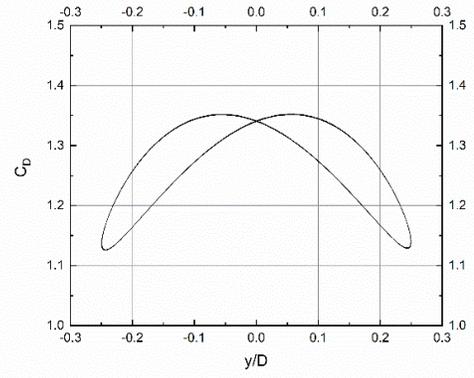
C4, drag coefficient

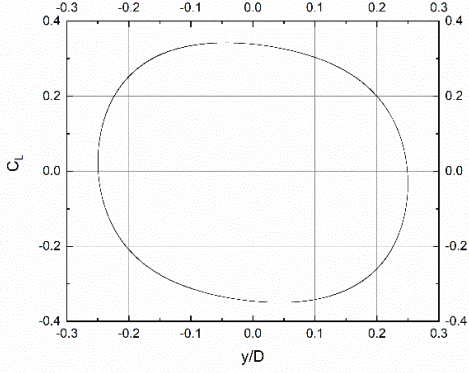
C5, lift coefficient

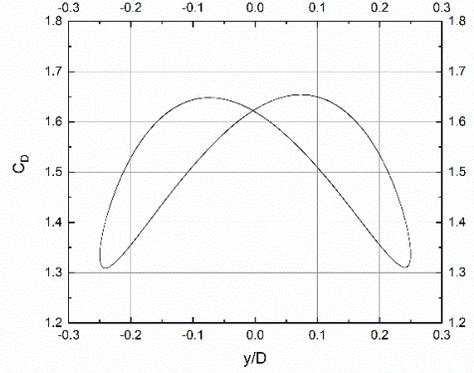
C5, drag coefficient

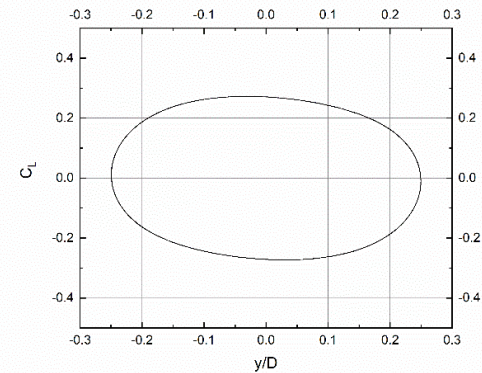
C6, lift coefficient

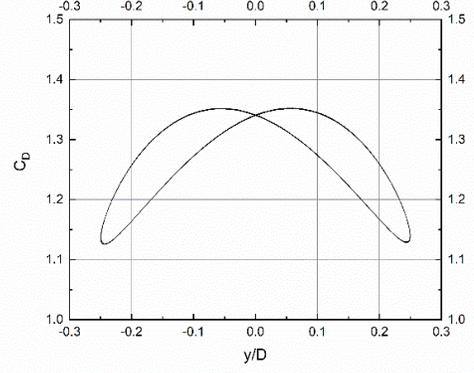
C6, drag coefficient

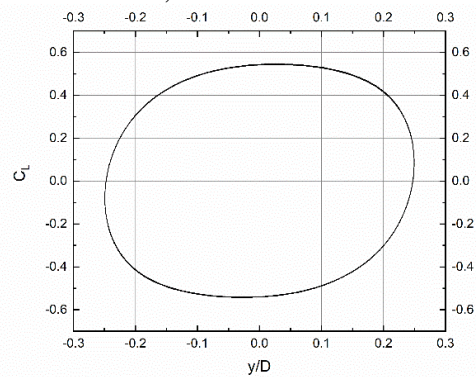
C7, lift coefficient

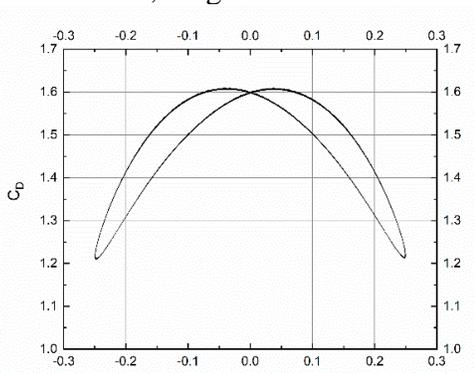
C7, drag coefficient

**Figure 5. Lift and Drag coefficients for cases C4, C5, C6 and C7**

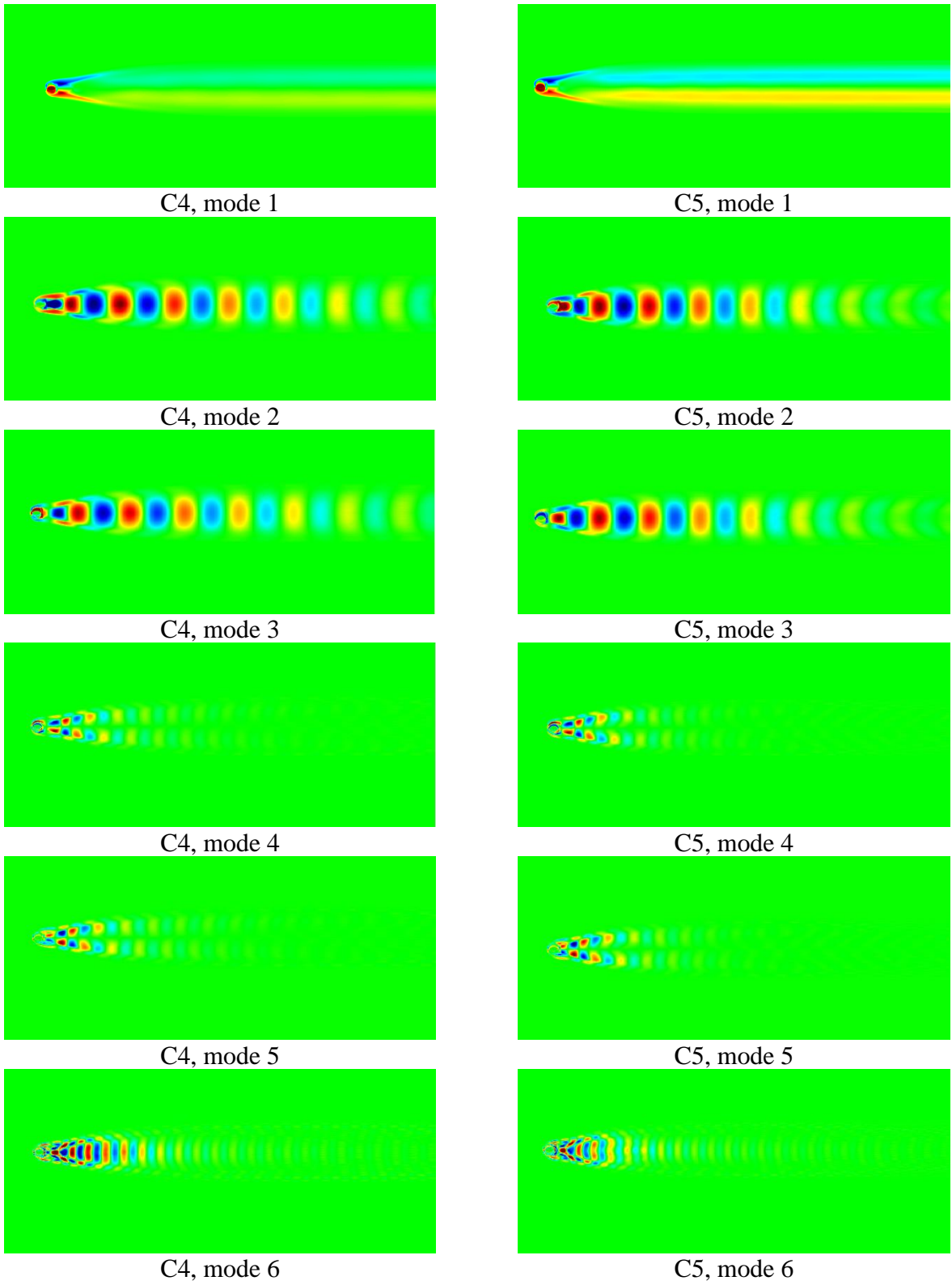

**Figure 6. POD modes for Cases C4 and C5**

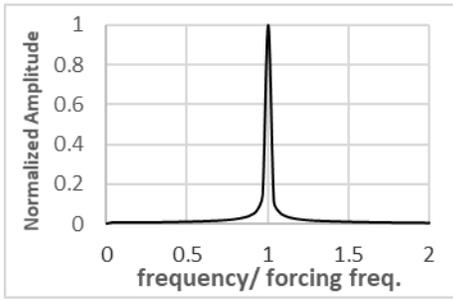
C4, mode 2

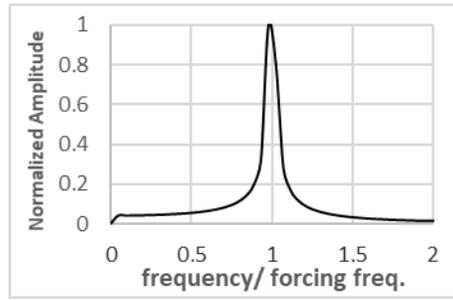
C5, mode 2

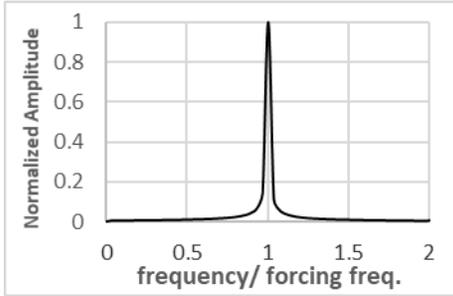
C4, mode 3

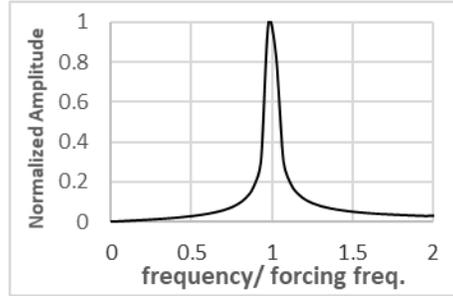
C5, mode 3

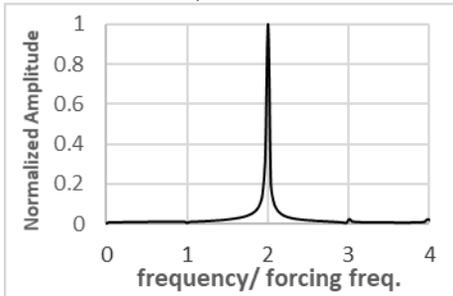
C4, mode 4

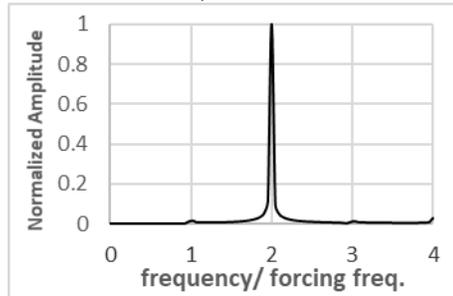
C5, mode 4

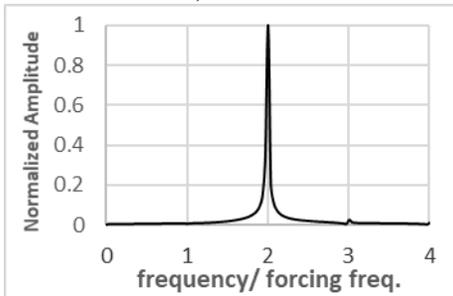
C4, mode 5

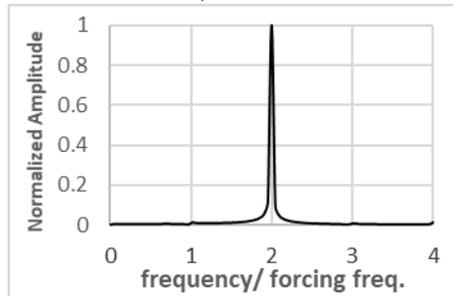
C5, mode 5

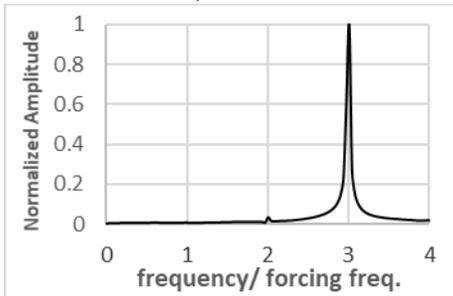
C4, mode 6

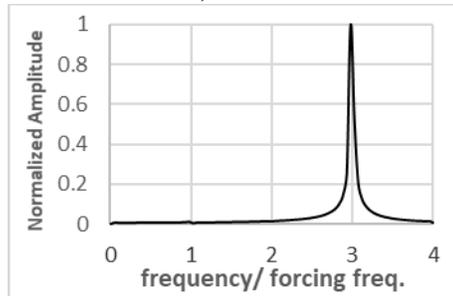
C5, mode 6

**Figure. 7. PSD plots for Cases C4 and C5**

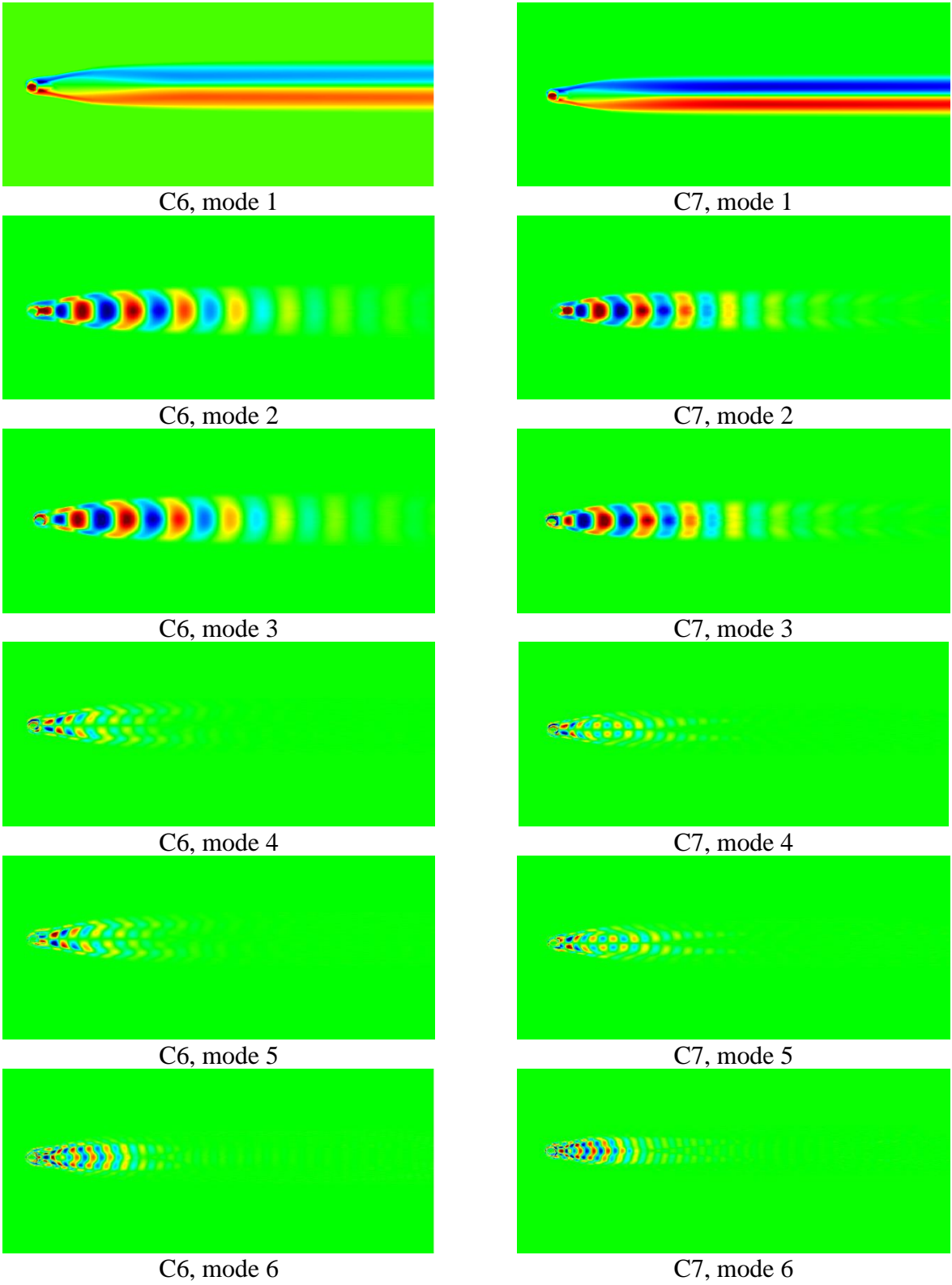

**Figure 8. POD modes for Cases C6 and C7**

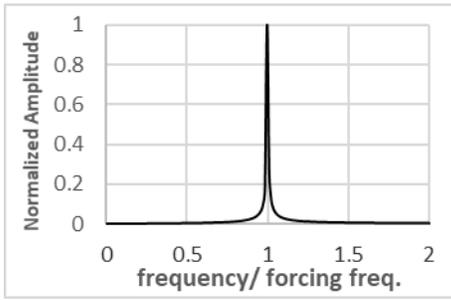
C6, mode 2
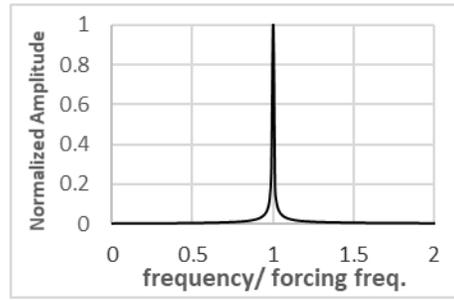
C7, mode 2
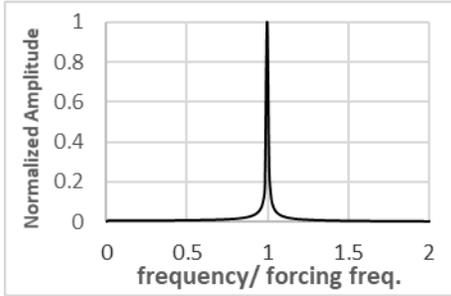
C6, mode 3
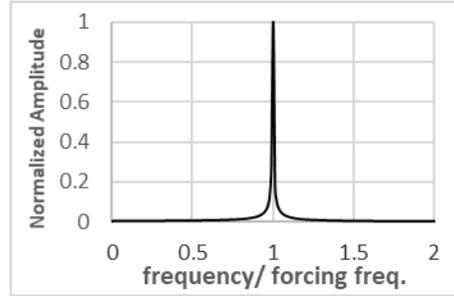
C7, mode 3
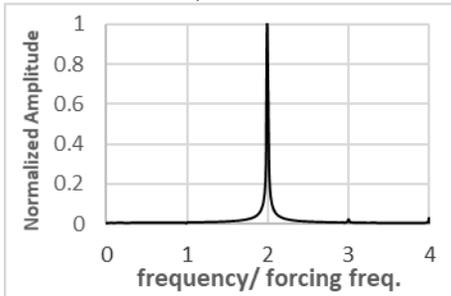
C6, mode 4
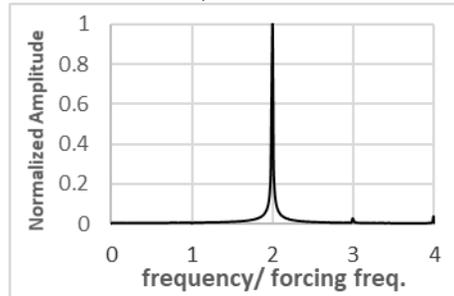
C7, mode 4
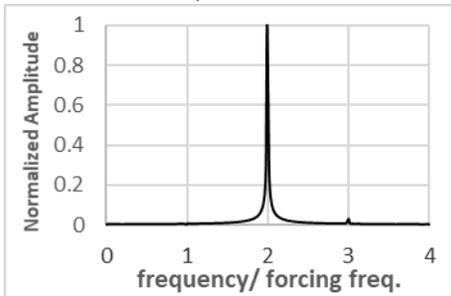
C6, mode 5
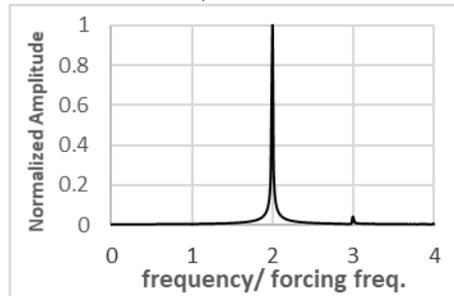
C7, mode 5
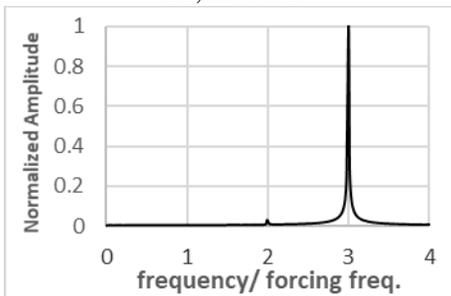
C6, mode 6
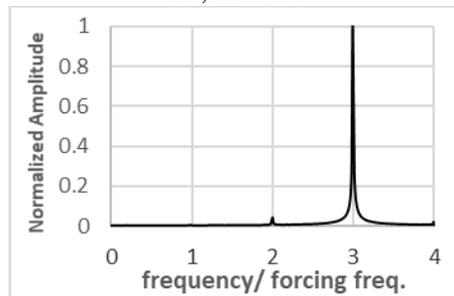
C7, mode 6

**Figure. 9. PSD plots for Cases C6 and C7**

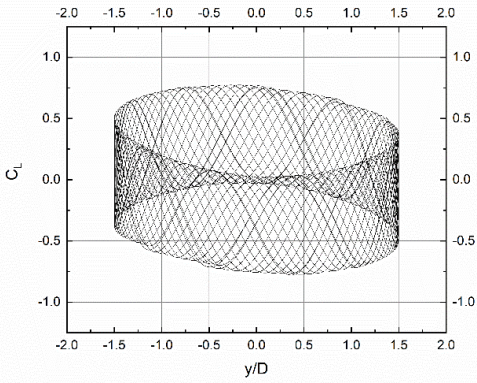
C8, lift coefficient

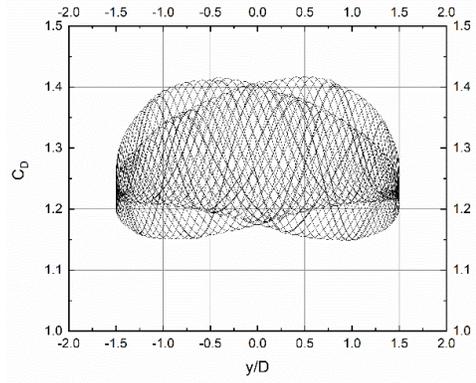
C8, drag coefficient

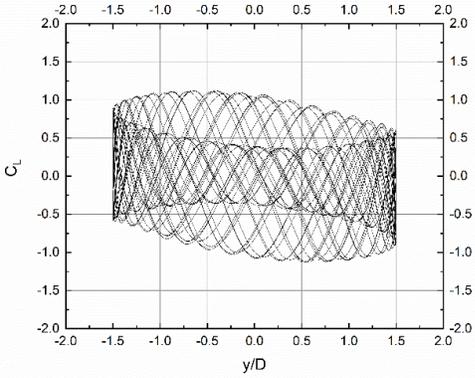
C9, lift coefficient

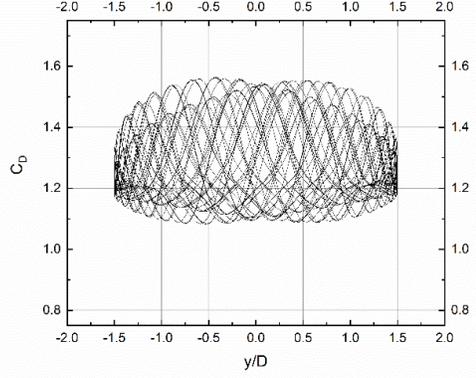
C9, drag coefficient

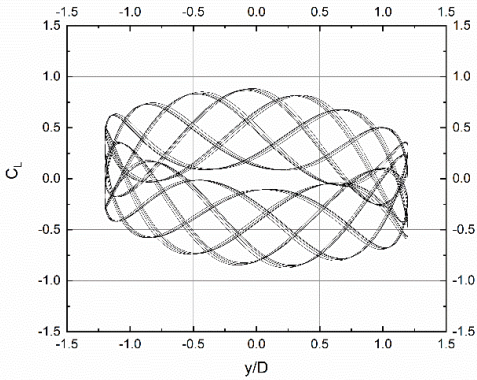
C10, lift coefficient

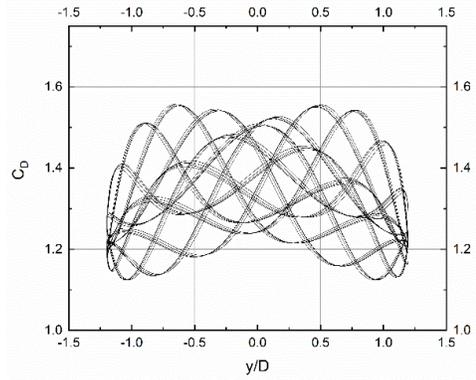
C10, drag coefficient

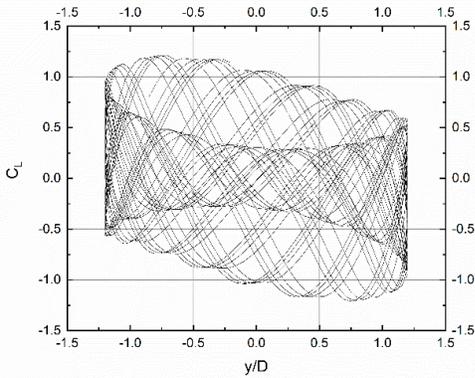
C11, lift coefficient

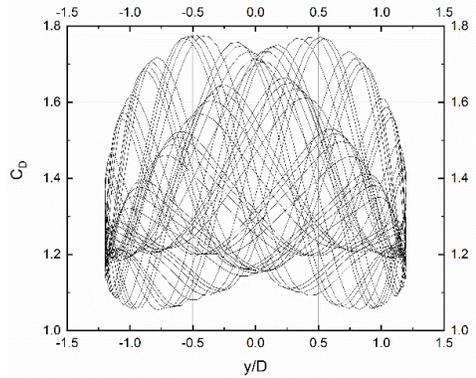
C11, drag coefficient

**Figure 10. Lift and Drag coefficients for cases C8, C9, C10 and C11**

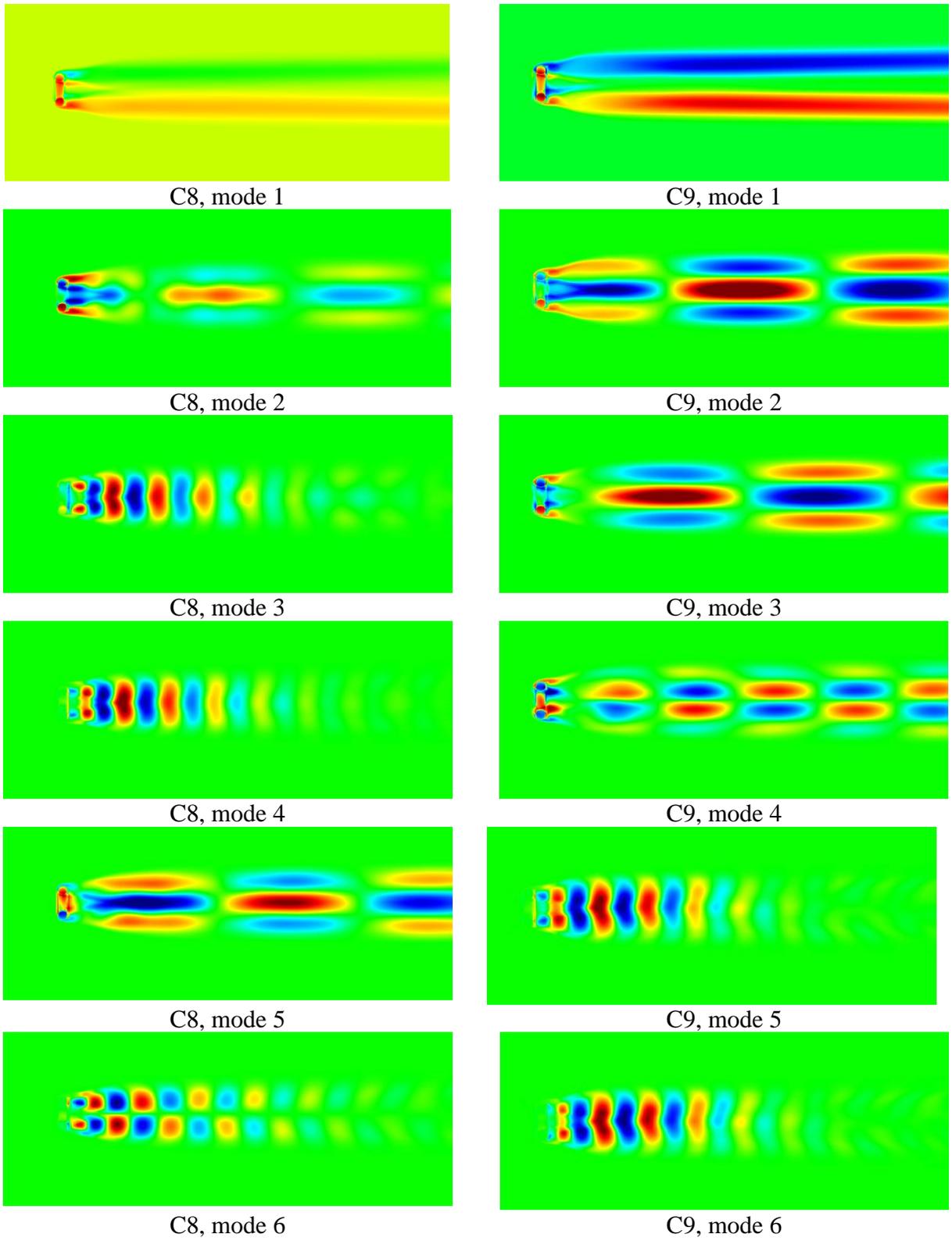

**Figure 11. POD modes for Cases C8 and C9**

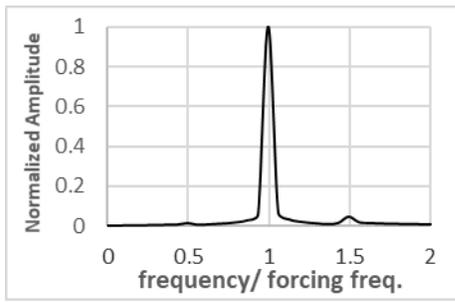
C8, mode 2

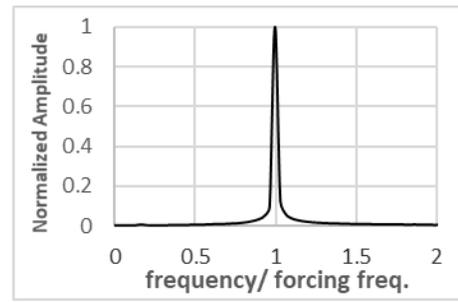
C9, mode 2

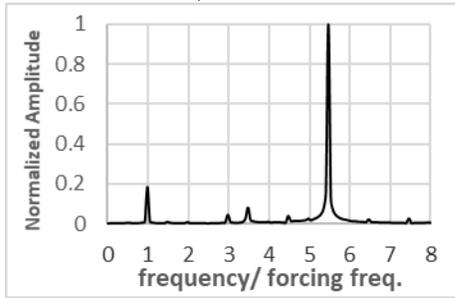
C8, mode 3

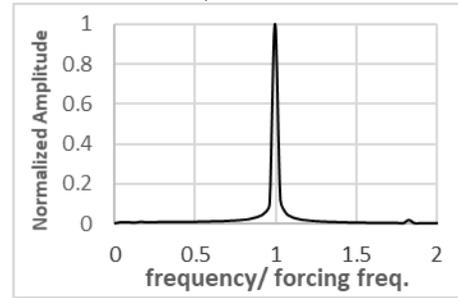
C9, mode 3

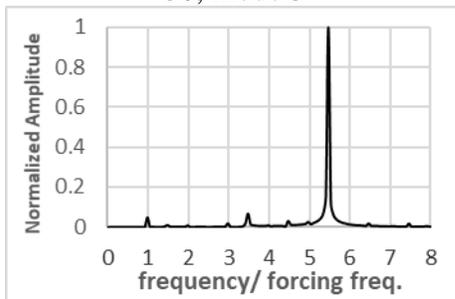
C8, mode 4

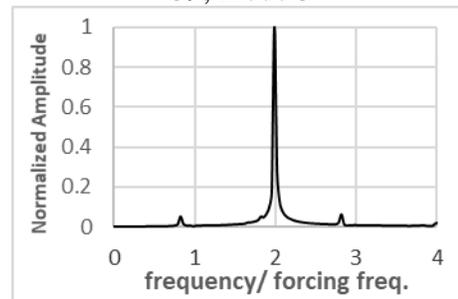
C9, mode 4

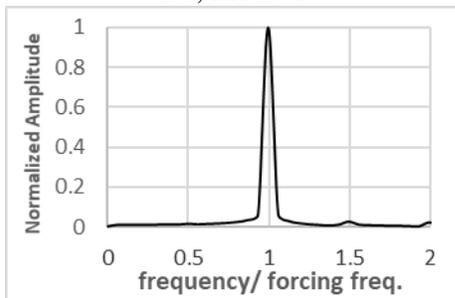
C8, mode 5

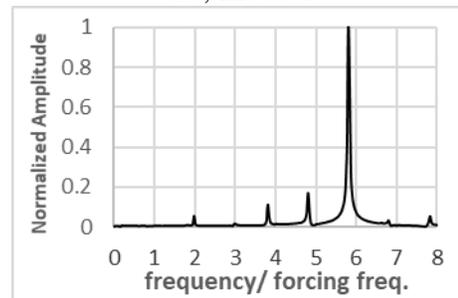
C9, mode 5

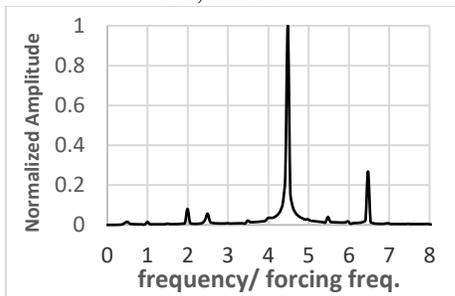
C8, mode 6

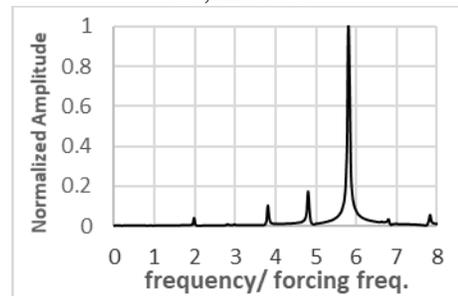
C9, mode 6

**Figure.12. PSD plots for Cases C8 and C9**

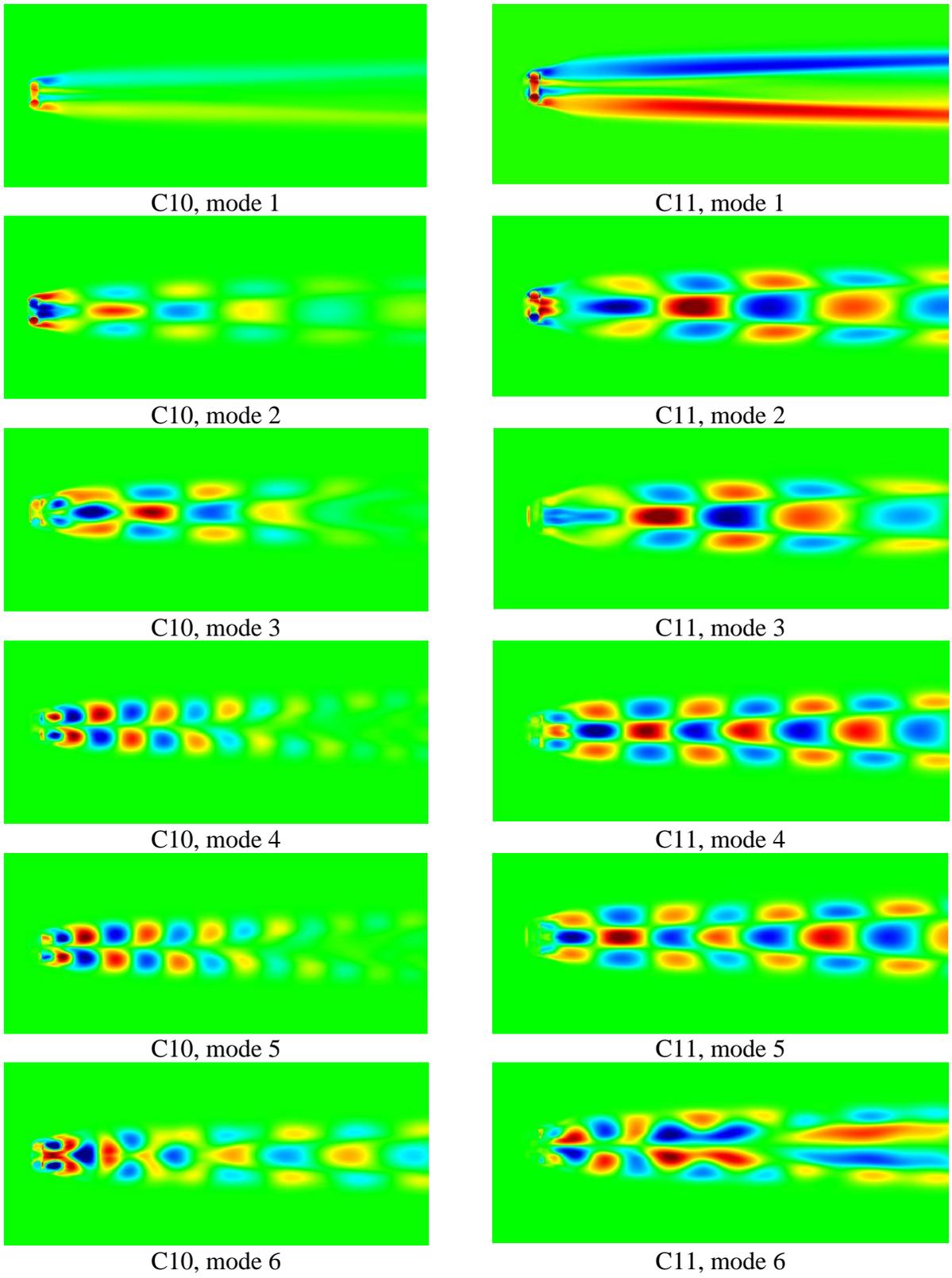

**Figure 13. POD modes for Cases C10 and C11**

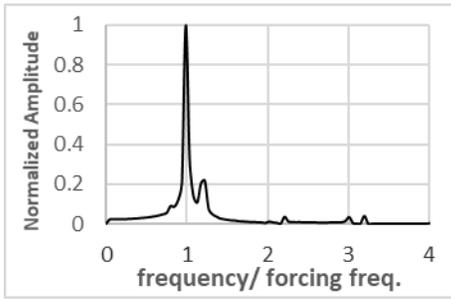
C10, mode 2

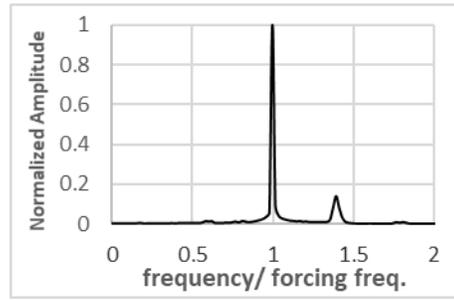
C11, mode 2

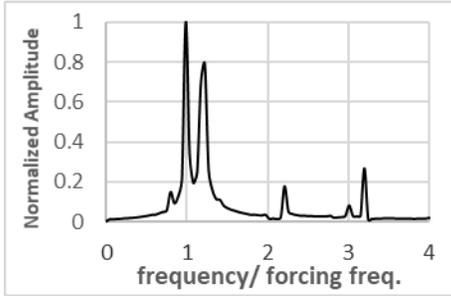
C10, mode 3

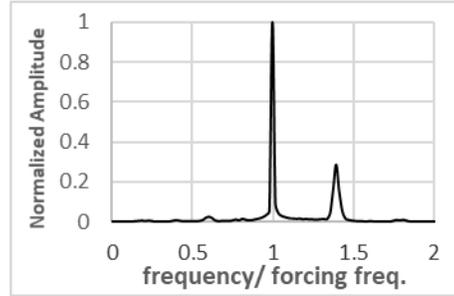
C11, mode 3

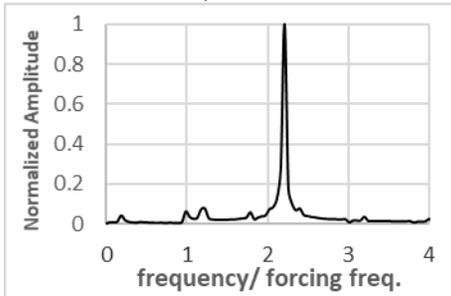
C10, mode 4

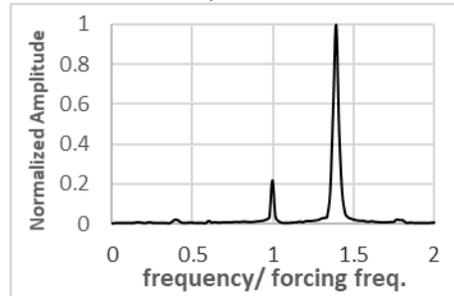
C11, mode 4

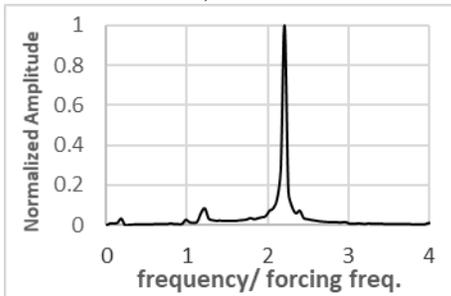
C10, mode 5

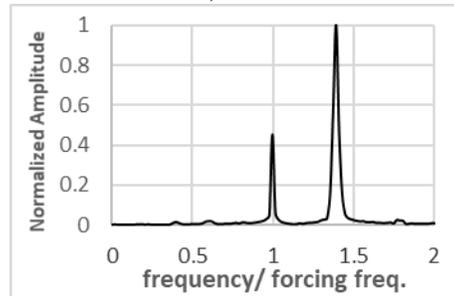
C11, mode 5

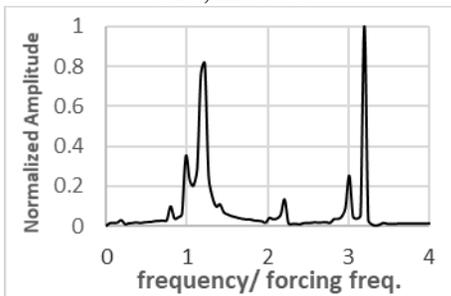
C10, mode 6

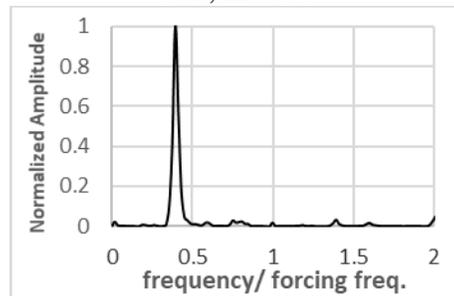
C11, mode 6

**Figure. 14. PSD plots for Cases C10 and C11**